\documentstyle[11pt,epsf]{article}
\begin{document}

\catcode`@=11
\long\def\@caption#1[#2]#3{\par\addcontentsline{\csname
  ext@#1\endcsname}{#1}{\protect\numberline{\csname
  the#1\endcsname}{\ignorespaces #2}}\begingroup
    \small
    \@parboxrestore
    \@makecaption{\csname fnum@#1\endcsname}{\ignorespaces #3}\par
  \endgroup}
\catcode`@=12
\newcommand{\newc}{\newcommand}
\newc{\gsim}{\lower.7ex\hbox{$\;\stackrel{\textstyle>}{\sim}\;$}}
\newc{\lsim}{\lower.7ex\hbox{$\;\stackrel{\textstyle<}{\sim}\;$}}
\newc{\gev}{\,{\rm GeV}}
\newc{\mev}{\,{\rm MeV}}
\newc{\ev}{\,{\rm eV}}
\newc{\kev}{\,{\rm keV}}
\newc{\tev}{\,{\rm TeV}}
\newc{\mz}{m_Z}
\newc{\mpl}{M_{Pl}}
\renewcommand{\phi}{\varphi}
\newc\order{{\cal O}}
\newc\CO{\order}
\newc\CL{{\cal L}}
\newc{\eps}{\epsilon}
\newc{\re}{\mbox{Re}\,}
\newc{\im}{\mbox{Im}\,}
\newc{\invpb}{\,\mbox{pb}^{-1}}
\newc{\invfb}{\,\mbox{fb}^{-1}}
%
%
\def\NPB#1#2#3{Nucl. Phys. {\bf B#1} (19#2) #3}
\def\PLB#1#2#3{Phys. Lett. {\bf B#1} (19#2) #3}
\def\PLBold#1#2#3{Phys. Lett. {\bf#1B} (19#2) #3}
\def\PRD#1#2#3{Phys. Rev. {\bf D#1} (19#2) #3}
\def\PRL#1#2#3{Phys. Rev. Lett. {\bf#1} (19#2) #3}
\def\PRT#1#2#3{Phys. Rep. {\bf#1} (19#2) #3}
\def\ARAA#1#2#3{Ann. Rev. Astron. Astrophys. {\bf#1} (19#2) #3}
\def\ARNP#1#2#3{Ann. Rev. Nucl. Part. Sci. {\bf#1} (19#2) #3}
\def\MPL#1#2#3{Mod. Phys. Lett. {\bf #1} (19#2) #3}
\def\ZPC#1#2#3{Zeit. f\"ur Physik {\bf C#1} (19#2) #3}
\def\APJ#1#2#3{Ap. J. {\bf #1} (19#2) #3}
\def\AP#1#2#3{{Ann. Phys. } {\bf #1} (19#2) #3}
\def\RMP#1#2#3{{Rev. Mod. Phys. } {\bf #1} (19#2) #3}
\def\CMP#1#2#3{{Comm. Math. Phys. } {\bf #1} (19#2) #3}
\relax
%
%
%
\def\beq{\begin{equation}}
\def\eeq{\end{equation}}
\def\bea{\begin{eqnarray}}
\def\eea{\end{eqnarray}}
%
%
%
\newc{\ie}{{\it i.e.}}          \newc{\etal}{{\it et al.}}
\newc{\eg}{{\it e.g.}}          \newc{\etc}{{\it etc.}}
\newc{\cf}{{\it c.f.}}
%
%
%
%
\def\slash#1{\rlap{$#1$}/} 
\def\Dsl{\,\raise.15ex\hbox{/}\mkern-13.5mu D} 
\def\delsl{\raise.15ex\hbox{/}\kern-.57em\partial}
\def\Ksl{\hbox{/\kern-.6000em\rm K}}
\def\Asl{\hbox{/\kern-.6500em \rm A}}
\def\Qsl{\hbox{/\kern-.6000em\rm Q}}
\def\gradsl{\hbox{/\kern-.6500em$\nabla$}}
%
%
%
\def\bar#1{\overline{#1}}
\def\vev#1{\left\langle #1 \right\rangle}
%

\begin{titlepage}
\begin{flushright}
{UCB--PTH--99/13\\
LBNL--43085\\
hep-ph/9904236\\
April 1999\\
}
\end{flushright}
\vskip 2cm
\begin{center}
{\large\bf Electroweak Symmetry Breaking and Large
Extra Dimensions}
\vskip 1cm
{\normalsize
Lawrence Hall and Christopher Kolda}\\
\vskip 0.5cm
{\it Department of Physics\\
University of California \\
Berkeley, CA~~94530, USA\\
and \\
Theory Group\\
Lawrence Berkeley National Laboratory\\
Berkeley, CA~~94530, USA\\}
\end{center}
\vskip .5cm
\begin{abstract}

If spacetime contains large compact extra dimensions,
the fundamental mass scale of nature, $\Lambda$,
may be close to the weak scale, allowing
gravitational physics to significantly modify electroweak symmetry breaking.
Operators of the form $(1 / \Lambda^2) |\phi^\dagger
D^\mu \phi|^2$ and $( 1 / \Lambda^2) \phi^\dagger W_{\mu\nu}B^{\mu\nu}
\phi$, where $W_{\mu\nu}$
and $B^{\mu\nu}$ are the $SU(2)$ and $U(1)$ field strengths and $\phi$
is the Higgs field, remove the
precision electroweak bound on the Higgs boson mass for values of $\Lambda$
in a wide range: $4 \tev \lsim \Lambda \lsim 11 \tev$.
Within this framework,
there is no preference between a light Higgs boson, a heavy Higgs boson, or a
non-linearly realized $SU(2) \times U(1)$ symmetry beneath
$\Lambda$. If there is a Higgs doublet, then
operators of the form $(1/\Lambda^2) \phi^\dagger
\phi (G^2, F^2)$, where $G_{\mu\nu}$ and
$F_{\mu\nu}$ are the QCD and electromagnetic field
strengths, modify the production of the Higgs boson by gluon-gluon fusion,
and the decay of the Higgs boson to $\gamma \gamma$, respectively. At Run
II of the Tevatron collider, a $\gamma \gamma$ signal for extra dimensions
will be discovered if $\Lambda$ is below 2.5 (1) TeV for a Higgs boson of
mass 100 (300) GeV. Furthermore, such a signal would point to
gravitational physics, rather than to new conventional gauge theories at
$\Lambda$. The discovery potential of the
LHC depends sensitively on whether the gravitational amplitudes interfere
constructively or destructively with the standard model amplitudes, and
ranges from $\Lambda = $ 3 -- 10 (2 -- 4) TeV for a light (heavy) Higgs boson.
\end{abstract}
\end{titlepage}
\setcounter{footnote}{0}
\setcounter{page}{1}
\setcounter{section}{0}
\setcounter{subsection}{0}
\setcounter{subsubsection}{0}


\section{Introduction} \label{sec:intro}

The conventional framework for particle physics beyond the standard
model (SM) assumes that the fundamental mass scale of nature is the Planck
mass: $\mpl \approx 10^{19}$ GeV. It is then natural to ask: why are the
masses of the
elementary particles so small? Proposed solutions to this
hierarchy problem have a common feature: new non-perturbative gauge
interactions dynamically generate a much lower scale, $M_{dyn}$, from which
electroweak symmetry breaking is generated, and hence all the masses
of the known elementary particles. Schematically, this mass hierarchy is
\begin{equation}
\mpl \rightarrow M_{dyn} \rightarrow M_W \,\,\ldots\,\, m_e.
\label{eq:masshier}
\end{equation}
In supersymmetric theories, $M_{dyn}$ is the scale at which
supersymmetry is broken, and the triggering of electroweak symmetry
breaking may be mediated, for example, by gravitational-scale physics,
or by gauge
interactions at much lower energy scales. Alternatively, $M_{dyn}$ may
be the scale of a new gauge force, technicolor, which forms fermion
condensates that directly break $SU(2) \times U(1)$. Finally, new strong
gauge forces could bind a composite Higgs boson.

Recently an alternative framework has been proposed~\cite{lowmpl} in which
spacetime is enlarged to contain large extra compact spatial dimensions.
At distances smaller than the size of these extra dimensions the
gravitational force varies more rapidly than the inverse square law, so
that the fundamental mass scale of gravity can be made much smaller
than $\mpl$. The conventional mass hierarchy of (\ref{eq:masshier})
is completely avoided if this fundamental mass scale is of order the weak
scale. In this case,
the length scale of the extra dimensions is much larger than the scales
probed experimentally at colliders, and hence this framework requires
that the quarks, leptons and gauge quanta of the SM are
spatially confined to a $3+1$ dimensional sub-space of the enlarged
spacetime.

The physics at the fundamental scale, $\Lambda$, which may well be that of
string theory, will be directly accessible to colliders of
sufficiently high energy; but even at lower energies this physics may
be experimentally probed.
At energies below the fundamental mass scale, physics is described by
an effective Lagrangian, which we take to be the most general set of
$SU(3) \times SU(2) \times U(1)$ invariant operators
involving quark, lepton and Higgs doublet fields of the SM:
\beq
\CL_{ef\!f} = \CL_{S\!M} + \sum_i {c_i \over \Lambda^p} \CO_i^{4+p}
\label{eq:leff}
\eeq
where $\CL_{S\!M}$ is the SM Lagrangian, $i$ runs over all
gauge invariant operators, $\CO_i^{4+p}$, of dimension $4+p$ with $p\geq1$, and
$c_i$ are unknown dimensionless couplings.

In this letter we study consequences of several of the dimension-6
operators. First we derive bounds on the $c_i/\Lambda^2$ from existing
experimental results under very conservative assumptions about
flavor-breaking in the ultraviolet theory. We then re-examine the
precision electroweak bounds on the Higgs boson mass. Analyses within
the standard model find a light Higgs; however, we will show that such
results do not survive the addition of non-renormalizable operators,
even if those operators are suppressed by scales as large as
$11\tev$.
In theories with large extra dimensions there is no good argument for
a light Higgs over a heavy Higgs or a non-linearly realized $SU(2)
\times U(1)$ symmetry, in which case (\ref{eq:leff}) must be replaced by a
chiral Lagrangian.
Finally we examine two operators in particular and their
effects on the discovery of Higgs bosons:
\bea
\CO_G &=& \phi^\dagger \phi\, G^a_{\mu\nu}G^{a\mu\nu}
\label{eq:hop1} \\
\CO_\gamma&=&\phi^\dagger \phi\, F_{\mu\nu}F^{\mu\nu}
\label{eq:hop2}
\eea
where $G^a_{\mu\nu}$ and $F^{\mu\nu}$
are the QCD and electromagnetic field strengths, and
$\phi$ is the Higgs doublet with Re$\,\phi^0=(v+h)/\sqrt{2}$.
The first operator contributes to Higgs production at hadron colliders via
gluon-gluon fusion, and the second to Higgs decay to $\gamma \gamma$.
There are two reasons why these effects provide a significant
discovery potential for extra dimensions: first, they are competing against a
SM signal which is suppressed by loop factors, and second, the
SM $\Gamma(h\to\gamma\gamma)$ is further suppressed by $e^4
\simeq 10^{-2}$, where $e$ is the electromagnetic coupling
constant.

However we assume that the physics at scale $\Lambda$ which
generates (\ref{eq:hop1})--(\ref{eq:hop2}),
does so in a way that the coefficients are
not suppressed by powers of the SM gauge coupling constants (see also
\cite{bdn}).
Such a behavior is certainly {\em not}\/ expected if the theory at
$\Lambda$ is a 4-dimensional gauge field theory: in that case operators of
the form (\ref{eq:hop1})--(\ref{eq:hop2}) would arise by integrating
out heavy fields, but these fields must couple to $F_{\mu\nu}$ and
$G_{\mu\nu}$ with the usual SM gauge couplings, and further, as shown
in $\cite{einhorn}$, they must be also be loop-suppressed. Thus even
if the gauge theory at $\Lambda$ were strongly-coupled, it seems
unlikely that coefficients of $\CO(1)$ could be generated.
This is very
important --- the effect of the interaction $(e^2/ \Lambda^2)
\phi^\dagger \phi F^2$ on the $h \rightarrow \gamma \gamma$ branching
ratio has been studied, and is small for $\Lambda \geq 1\tev$~\cite{hagiwara}.
Thus observation of the physics we will describe in
Section~\ref{sec:expt} would provide support for
an extra-dimensional theory.

\section{Some Constraints on $\Lambda$}

Are the coefficients $c_{G, \gamma}/ \Lambda^2$ expected to be large
enough for an observable $h \rightarrow \gamma \gamma$ signal? In
general this cannot be excluded, since physics induced by operators
$\CO_i$ will place bounds on
\beq
\frac{f_i}{\Lambda_i^p} \equiv \frac{c_i}{\Lambda_{\phantom{i}}^p}
\quad\quad\quad (f_i=\pm1)
\eeq
not on $\Lambda$. However, it would be unreasonable to expect  $c_{G,
\gamma}$ to be orders of magnitude larger than all the other $c_i$.

It is tempting to assume that although the dimensionless coefficients
$c_i$ are unknown, they are all of order unity. However, in this case
operators which violate baryon number constrain $\Lambda \gsim 10^{16}
\gev$, and $CP$ violating operators contributing to $\epsilon_K$
constrain $\Lambda \gsim 10^{5}
\gev$. Thus the framework of large compact extra dimensions, allowing a
fundamental scale close to the weak scale, is clearly excluded unless
the low energy effective theory possesses an approximate flavor
symmetry, in which case one expects
\beq
c_i = \varepsilon_{Fi} \; c_i'
\label{eq:cprime}
\eeq
with $c_i'$ of order unity. The flavor symmetry breaking parameters,
$\varepsilon_{Fi}$, depend on the flavor symmetry group and the pattern
of flavor symmetry breaking. For operators which violate flavor and $CP$
they must be small, while for operators which conserve flavor and $CP$
they may be set to unity.

To allow low values for $\Lambda$, the flavor group should be large,
and its breaking should be kept to a minimum, consistent with the
observed quark and lepton masses and mixings. The maximum flavor
group of the SM is $U(3)^5$. The three
generations of quarks and leptons transform as $q_L = (u_L, d_L) \sim
(3,1,1,1,1)$; $u_R \sim (1,3,1,1,1)$; $d_R \sim (1,1,3,1,1)$;
 $\ell_L = (\nu_L, e_L) \sim (1,1,1,3,1)$; $e_R \sim (1,1,1,1,3)$. If
there are only three symmetry breaking parameters, one for each of the
up, down and charged lepton mass matrices, $\varepsilon_u
\sim(3,\bar{3},1,1,1)$; $\varepsilon_d \sim(3,1,\bar{3},1,1)$; $\varepsilon_e
\sim(1,1,1,3,\bar{3})$, then baryon number and
lepton number remain unbroken. (The $\varepsilon_i$ are equal to the
Yukawa couplings up to an $\CO(1)$ factor, $c_i$:
$\lambda_{u,d,e}=c_{u,d,e}\varepsilon_{u,d,e}$.)
However, even after imposing such a
flavor symmetry, there remain operators such as
\beq
\CO_{qq} = (\bar{q}_L \gamma^\mu \varepsilon_u \varepsilon_u^\dagger q_L)^2
=c_u^4(\bar q_L\gamma^\mu \lambda_u\lambda_u^\dagger q_L)^2
\label{eq:oqq}
\eeq
which contribute to $\epsilon_K$ and
constrain $\Lambda \gsim 4.2 \tev\times (\sqrt{c_{qq}}/c_u^2) $. There are
two ways to avoid this bound. First, since the bound depends
quadratically on $c_u$, values slightly larger than 1 will
weaken the bound significantly;
this seems entirely natural to us. Second, one could
postulate that $\varepsilon_{u,d}$ are real and the observed
$\epsilon_K$ has an exotic origin; we view this as disfavored
given that measurements of $V_{ub}/V_{cb}$ and $B-\bar B$ mixing
indicate values of the CKM matrix elements consistent with a standard
model origin of $\epsilon_K$ to better than $30\%$.

For the $h \rightarrow \gamma \gamma$ signal, we are interested in the
operators (\ref{eq:hop1})--(\ref{eq:hop2}),
which conserve $U(3)^5$. Hence,
even if the higher dimension flavor violating operators, such as
(\ref{eq:oqq}), are completely absent, it is important to study
constraints on $\Lambda$ expected from operators which conserve
$U(3)^5$. Such operators include
flavor-conserving four-fermion operators and operators involving the
Higgs doublet and the gauge fields.
There have been many analyses to date which obtain
constraints from these operators, and here we will simply repeat
the results of these analyses, in the notation we are using for
$\Lambda$.
(An analysis similar to ours was recently presented in~\cite{bdn}.)

Among the $CP$-conserving
four-fermion operators, the strongest constraints come from
atomic parity violation. The operator
\begin{equation}
\CO_{\ell q} = (\bar \ell_L\gamma_\mu \ell_L)(\bar q_L\gamma^\mu
q_L)
\label{eq:olq}
\end{equation}
gives a constraint $\Lambda_{lq} >3.0\tev$~\cite{apv} at 95\% CL.
If the operator $(\bar e_R\gamma_\mu e_R)(\bar q_L\gamma^\mu
q_L)$ were generated with the same coefficient, $P$ would be preserved
in atomic systems and the previous limit would vanish.
Although we do not expect $P$ to be a good symmetry of the underlying
theory, a partial cancellation could easily weaken this bound.
Apart from $P$-violation, the best bounds on $\Lambda_{\ell q}$
currently come from  OPAL~\cite{eebb},
using the $\ell_1, q_3$ component, and from CDF~\cite{mumuqq}, using
the $\ell_2, q_1$ component. Both find $\Lambda>800\gev$ at
95\% CL.

The bounds on the coefficients of the operators $\CO_{qq, \ell q}$ of
(\ref{eq:oqq})--(\ref{eq:olq}) do not provide strict bounds on the scale
$\Lambda$, because $\Lambda = \Lambda_i \sqrt{c_i}$, and the $c_i$ are
unknown. Nevertheless, if the (flavor-conserving)
$c_i'=c_i$ are expected to be of order unity for
these operators, then $\Lambda \gsim 3 \tev$ is clearly allowed, while a
value of $\Lambda$ as low as $1 \tev$ seems disfavored.

\section{Precision Electroweak Physics and the Higgs Mass Bound}

A second class of constraints arise from precision measurements in the
electroweak gauge sector, namely from the $S$ and $T$ parameters (see,
\eg, \cite{hagiold}). The
strongest of these constraints arise from the operators:
\bea
\CO_{BW}&=& B^{\mu \nu} (\phi^\dagger \tau^a
W^{a\mu\nu} \phi) \label{obw} \\
\CO_{\Phi}&=& (\phi^\dagger D^\mu\phi)
(D_\mu\phi^\dagger\phi) \label{op1}
\eea
which contribute
\bea
\Delta S_{new} &=& -\frac{2c_Ws_W}{\alpha}\frac{v^2}{\Lambda_{BW}^2}f_{BW} \\
\Delta T_{new} &=&
-\frac{1}{2\alpha}\frac{v^2}{\Lambda_{\Phi}^2}f_{\Phi}
\eea
where $s_W,c_W$ are the sine and cosine of the weak angle and
$f_{BW}$, $f_\Phi$ are unknown signs.

A global fit to electroweak observables~\cite{erler} yields
$S_{f\!it}=-0.14\pm0.12$ and $T_{f\!it}=-0.22\pm0.15$ assuming
$m_h=100\gev$.~\footnote{The fit in \cite{erler} uses $m_h=600\gev$ and 
defines $S=T=0$ in the SM. We rescale to $m_h=100\gev$ using the
parameterization of Ref.~\cite{hagiold} (see
Eqs.~(\ref{dS})--(\ref{dT})). We then treat deviations from
$m_h=100\gev$ as ``new physics.''} 
Since each operator contributes only to one of $S$ or $T$, we can find
independent bounds on each. We find that at 95\% CL:
\bea
\Lambda_{BW}&>& 3.6\tev \label{bwbound}\\
\Lambda_{\Phi}&>& 3.0 \tev.
\eea
We can also extract a bound if
$\Lambda_{BW}=\Lambda_{\Phi}$:
$\Lambda > 4.0\tev$,
allowing the Higgs mass to vary over the range
$100\gev<m_h<800\gev$. We see that the constraints from precision
electroweak physics are very similar in magnitude to those obtained in
the previous section.

How important are these constraints for restricting $\Lambda_\gamma$?
Although the electromagnetic field strength, $F^{\mu \nu}$, is not $SU(2)\times
U(1)$ invariant, the operator $\CO_\gamma$ is generated, after
electroweak symmetry-breaking, from the
invariant operators $\CO_B=
(\phi^\dagger\phi)B_{\mu\nu}B^{\mu\nu}$,
$\CO_W=(\phi^\dagger\phi)W_{\mu\nu}W^{\mu\nu}$ and $\CO_{BW}$ of
Eq.~(\ref{obw}):
\beq
\frac{f_\gamma}{\Lambda_\gamma^2}=
c_W^2\frac{f_{B}}{\Lambda_{B}^2}+s_W^2 \frac{f_{W}}{\Lambda_{W}^2}+
c_Ws_W\frac{f_{BW}}{\Lambda_{BW}^2}.
\label{foper}
\eeq
{\em If}\/ all $f_i$ and $\Lambda_i$ on the right side of Eq.~(\ref{foper})
were equal, then the bound (\ref{bwbound})
on $\Lambda_{BW}$ implies
$\Lambda_\gamma > 3.3\tev$. However, changes in the relative signs
or sizes of each contribution significantly reduces the bound;
thus we have no strong lower bound on the scale $\Lambda_\gamma$ itself.
Likewise we know of no strong constraint on the scale $\Lambda_G$ either.

Finally we wish to address the question of the Higgs mass. It is
well-known that fits to the electroweak data indicate a light Higgs. A
simple fit can be done using only $S$ and $T$ as given above and the
following parameterization of the Higgs contributions from
Ref.~\cite{hagiold}:
\bea
\Delta S_H&=&0.091 x_H - 0.010 x_H^2 \label{dS}\\
\Delta T_H&=&-0.079 x_H-0.028 x_H^2+0.0026 x_H^3 \label{dT}
\eea
where $x_H=\log(m_h/100\gev)$. Using these forms, one can do a fit demanding
$S_{f\!it}=\Delta S_H+\Delta S_{new}$ and likewise for $T$.
For the SM alone, a 95\% CL upper bound of $255\gev$ has been
obtained~\cite{erler}. However it is clear that from the point of view
of the oblique parameters, shifts in $\Delta S_H$ and $\Delta T_H$ can
be compensated by similar shifts in $\Delta S_{new}$ and $\Delta
T_{new}$. Thus we can derive an effective
``95\% CL bound'' on the Higgs mass as
a function of $\Lambda$ under the requirement that the fit to the
experimentally obtained $S_{f\!it}$ and $T_{f\!it}$ be no worse than
that obtained for $m_h=255\gev$ and $\Lambda\to\infty$. (We do this by
constructing a $\chi^2$ distribution from $S$ and $T$ alone.)

How large can the Higgs mass become with
the inclusion of $\CO_{BW}$ and $\CO_{\Phi}$? The answer is: quite large.
Fitting to $m_h$ as a function of $\Lambda$ and using $S$ and $T$ as
``experimental'' inputs, we find
for particular choices of the signs of the operators (\ie, $f_{BW}=f_\Phi=+1$)
that {\em the precision electroweak bound on the Higgs mass disappears
completely for $4\tev\lsim\Lambda\lsim 11\tev$!} (By ``disappear'' we
mean that the 95\% upper bound on $m_h$ exceeds the unitarity bound of
approximately $800\gev$ and so is meaningless.) Thus, in the context
of gravitational physics at or below $10\tev$, the usual claims that
electroweak physics prefers
a light Higgs do not hold. And even for $\Lambda$ as high as $17\tev$,
the upper limit on the Higgs mass exceeds $500\gev$.

\begin{figure}[t]
\centering
\epsfxsize=4truein
\epsffile{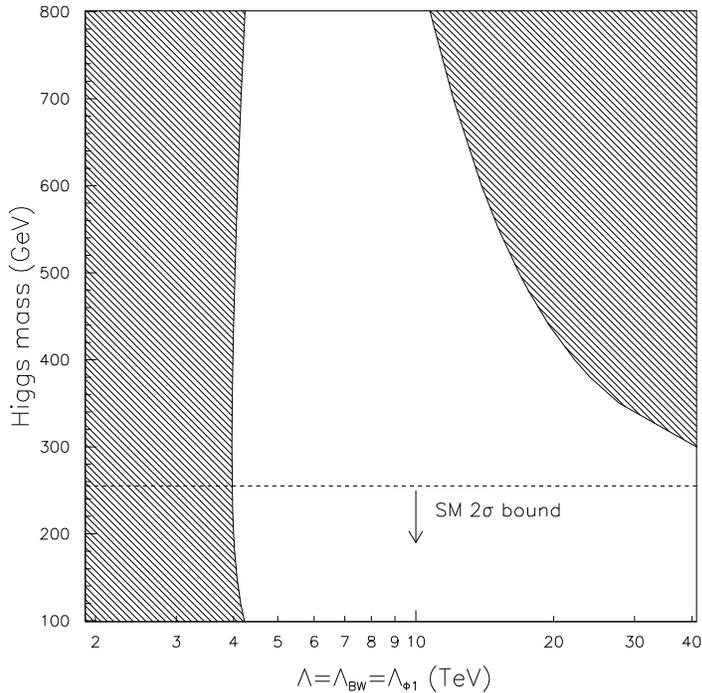}
\caption{Precision electroweak limits on the Higgs mass as a function
  of the scale of new physics. For this figure, $\Lambda_{BW}$ and
  $\Lambda_{\Phi}$ are chosen equal, while the signs $f_{BW}$ and
  $f_\Phi$ are chosen to
  maximize the allowed region. Hatched regions are disallowed at
  95\%, while the dashed line borders the region allowed in the
  SM alone.}
\label{massfig}
\end{figure}

These results are
summarized in Figure~\ref{massfig} where we show the 95\% CL allowed
range for $m_h$ as a function of $\Lambda\equiv\Lambda_{BW}=\Lambda_{\Phi}$.
The hatched region at small $\Lambda$ is ruled out because of its large
contribution to $S$ and $T$, while the region at large $\Lambda$ and
large $m_h$ is ruled out because the new operators contribute too
little to $S$ and $T$ to significantly effect the SM fit to the Higgs
mass. However for intermediate $\Lambda$ (unhatched region)
it is clear that there is
effectively no limit on the Higgs mass thanks to the effects of the
new operators.

(If the physics at $\Lambda$ were weakly-coupled then we would expect
that $c_{BW}\simeq e^2c_Ws_W$; then allowing $c_\Phi\simeq1/4$ would
reproduce Fig.~\ref{massfig}, only with the $\Lambda$ rescaled by
$\sim1/2$. Thus the preference for a light Higgs in the SM is even
removed for a weakly-coupled gauge theory if $\Lambda\sim 2-5\tev$.)

Finally, we note that the one other argument for a light Higgs, namely
triviality, is no longer applicable in these models either. With such
a low ultraviolet cutoff ($\Lambda\sim$ few TeV), the Higgs
self-coupling cannot run to its Landau pole for $m_h\lsim 1\tev$.

\section{Implications for Electroweak Symmetry Breaking}

The mechanism for electroweak symmetry breaking (EWSB) is
unknown. Nevertheless, it is commonly believed that the Higgs boson
exists, and is light. The two indirect indications for this are:
\begin{itemize}
\item The successful prediction of the weak mixing angle from gauge
coupling constant unification. This prediction results in theories
with weak scale supersymmetry which are perturbative to a high scale;
such theories have a light Higgs boson, $m_h \lsim 150 \gev$~\cite{kkw}.
\item The experimental values of the precision electroweak observables
are consistent with the standard model, at 95\% C.L., only if 
$m_h \lsim255 \gev$~\cite{erler}.
\end{itemize}

If there are large extra dimensions allowing the fundamental scale,
$\Lambda$, to be in the TeV domain, neither of these points can be used
to argue that the  Higgs boson is light. For the first: it has not
been demonstrated that it is
possible to predict the weak mixing angle to the percent level of
accuracy in these theories; furthermore, there is no need for the field theory
below $\Lambda$ to be supersymmetric
since there is no large hierarchy between the weak scale and $\Lambda$.
The argument from fits to the precision electroweak observables
applies only if the standard model is the correct theory up to scales
of at least 10 TeV; it is a very weak bound which is immediately
evaded by large extra dimensions, allowing several scenarios
for EWSB:
\begin{itemize}
\item {\em Light Higgs} ($m_h < 200 \gev$):  For $\Lambda \gsim 20 \tev$
some protection mechanism for the Higgs mass would be required; if
this is supersymmetry, the Higgs will be light.
For $ \Lambda \approx 1 - 3 \tev$, if the tree
level Higgs mass happened to vanish, EWSB and a light Higgs boson
could result from 1 loop radiative corrections.
\item {\em Heavy Higgs} ($m_h > 200 \gev$):  This could arise for
$\Lambda \approx 1 - 3$ TeV, if the Higgs mass parameter is somewhat less
than $\Lambda$, or alternatively for $\Lambda \approx 3 - 10 \tev$ if
the Higgs mass parameter vanishes at tree level but arises at 1
loop. In both cases a large value for the Higgs self coupling is
needed, and the operators (\ref{obw}) and (\ref{op1}) must mimic
the effects of a light Higgs in the $S$ and $T$ parameters.
\item{\em No Higgs}:  Physics at the fundamental scale $\Lambda \approx
1 - 3 \tev$ may itself cause EWSB. An example of this has already
been proposed \cite{ad}. In this case the theory below $\Lambda$
will have $SU(2) \times U(1)$ realized non-linearly, and the chiral
Lagrangian will
have operators analagous to (\ref{obw}) and (\ref{op1}) which mimic
the effects of a light Higgs in the $S$ and $T$ parameters.
\end{itemize}
A light Higgs boson is just one
possibility amongst several for EWSB, and is not
preferred.

We have shown that, in theories with large extra dimensions having 
$\CO_{BW,\Phi}$ with $c_{BW,\Phi}$ of order unity, the precision
electroweak data provide a lower bound on the fundamental scale,
$\Lambda_{min}\approx3\tev$. For values of $\Lambda$ in the range
(1--3)$\times\Lambda_{min}$, the signs $f_{BW,\Phi}$ are critical. For 
two sign choices, no successful fit can be found for any Higgs
mass. For a third choice, a good fit to the data is found for Higgs
masses all the way up to $m_h=800\gev$. For the final choice, masses
up to $800\gev$ are also obtained, though the fits are less convincing.
Only in the case of very large $\Lambda$ does the data still prefer a
light Higgs, but then the quadratic finetuning of the light Higgs mass 
to one part in $m_h^2/\Lambda^2$ is reintroduced.

In view of the bounds on $\Lambda_{min}$ of $3 - 4 \tev$ from each of
$\CO_{qq}$ (\ref{eq:oqq}),$\CO_{\ell q}$ (\ref{eq:olq}), $\CO_{BW}$
(\ref{obw}), and $\CO_{\Phi}$ (\ref{op1}), it may be felt that the
exciting possibility of $\Lambda$ in the $1 - 3 \tev$ range is
unlikely. Why would all the relevant $c_i$ coefficients be small? One
possibility is that the dominant interactions of the new physics at
$\Lambda$ preserve symmetries that are broken by the electroweak gauge
interactions, including $P$, $CP$ and custodial $SU(2)$. If these
symmetries are broken by sub-dominant interactions at $\Lambda$,
then the smallness of the relevant $c_i$ can be naturally explained.

\section{Higgs Production and Decay} \label{sec:expt}

For the case that there is a Higgs boson, either light or heavy,
we now study the effects of $\CO_{G,\gamma}$ of
(\ref{eq:hop1})--(\ref{eq:hop2}) on the
signal for $h \rightarrow \gamma \gamma$ at hadron colliders.
These operators have two immediate consequences.
First, when both Higgs fields are set to their vacuum expectation
values (vev's), the gauge
couplings of QED and QCD are shifted. But these shifts can be reabsorbed
into the definition of the gauge couplings and therefore have no
observable implications. (If one attempts to unify the SM gauge
couplings at some ultraviolet scale, or otherwise define theoretical
relations among them, then these shifts will enter into the relation
between the theoretical couplings and those extracted from
data. However, for all but the lightest $\Lambda$, this
shift is smaller than the experimental uncertainties.)

The second consequence is the possibility of unusual production and decay modes
of the (physical) Higgs bosons. Taking one of the Higgs fields to its vev, one
obtains terms in the effective Lagrangian:
\bea
\CL_{ef\!f} = \cdots +
 f_\gamma \frac{v}{\Lambda_\gamma^2}\,h\,F_{\mu\nu}F^{\mu\nu} +
 f_G \frac{v}{\Lambda_G^2}\,h\,G^a_{\mu\nu}G^{a\mu\nu} +\cdots
\label{eq:hleff}
\eea
where $h$ is the physical Higgs boson,
$v=246\gev$ and $f_{\gamma,g}= \pm1$ are unknown signs.
First, $\CO_G$ can contribute to the gluon fusion process $gg\to
h$. It is well-known that the dominant production mode for Higgs
bosons at the Tevatron and the LHC is
through gluon fusion, via a loop of $t$-quarks. Because the process
occurs at one-loop, non-renormalizable operators are more likely to provide a
significant correction to the cross-section. Integrating out the
$t$-quark, the relevant low-energy operator is then (for a recent discussion of
the relevant SM Higgs physics, see~\cite{nlo}):
\beq
\CL_{G,ef\!f}=\left(-\frac{g\alpha_s}{24\pi
    M_W}I_G+f_G\frac{v}{\Lambda_G^2} \right)\, h\,G^a_{\mu\nu}G^{a\mu\nu}
\eeq
where $g$ is the SU(2) coupling constant and
$I_G\to1(0)$ for $m_t^2\gg m_h^2$ $(m_t^2\ll m_h^2)$.
For $\Lambda\lsim4.5\tev$, the new
physics will actually dominate the production of Higgs bosons. Note that
the cross-section is maximized for constructive interference, $f_G=-1$,
and minimized for $f_G=+1$.

The operator $\CO_\gamma$ does not contribute to Higgs
production\footnote{However, a large coefficient to $\CO_\gamma$ could turn
the NLC into an $s$-channel Higgs factory when run in $\gamma\gamma$ mode.}.
However it can contribute to the decay of the Higgs into photons:
\beq
\Gamma(h\to\gamma\gamma)=\frac{|\beta|^2 m_h^3}{4\pi}
\eeq
for $\CL=\beta h F_{\mu\nu}F^{\mu\nu}$.
In the SM, this process is dominated by loops of
$W$-bosons and $t$-quarks. Integrating them out yields an effective
operator:
\beq
\CL_{\gamma,ef\!f}=\left(-\frac{g\alpha}{4\pi M_W}I_\gamma+
f_\gamma\frac{v}{\Lambda_\gamma^2}\right)\, h\,F_{\mu\nu}F^{\mu\nu}
\eeq
where $I_\gamma$ varies from roughly $-0.5$ to $-1.3$ as $m_h$ is
varied. Once again, the new physics will dominate the width
for $h\to\gamma\gamma$ given $\Lambda_\gamma\lsim 7\tev$. If $m_h\lsim
150\gev$, its decay width is dominated by final state $b$-quarks; then
$h\to\gamma\gamma$ becomes the dominant decay mode given
$\Lambda_\gamma\lsim 1.5\tev$. However, even for larger
$\Lambda_\gamma$, the branching ratio $h\to\gamma\gamma$ may be more
than sufficient to provide a strong signal. The signal is maximized
for $f_\gamma=+1$ (\ie, constructive interference of the SM and new physics)
and minimized for $f_\gamma=-1$.

(In the context of LEP, Ref.~\cite{eboli} recently examined the effect of
$\CO_\gamma$ and related operators on $e^+e^-\to 3\gamma, qq\gamma\gamma$ and
found sensitivity there to new physics roughly below a scale $\Lambda\lsim
600\gev$.)

Unfortunately, the operator $\CO_G$ can also contribute to the Higgs
decay width via $h\to gg$ which is unobservable among the QCD backgrounds.
In fact, to lowest order,
\beq
\Gamma(h\to gg)=8\left(\frac{\Lambda_\gamma}{\Lambda_G}\right)^4
\Gamma(h\to\gamma\gamma).
\eeq
In the limit in which the new physics is dominating the Higgs decays
and $\Lambda_\gamma\simeq\Lambda_G$, the $h\to gg$ decays suppress the
branching ratio into $h\to\gamma\gamma$ by about a factor of
10. However, once final state $WW/ZZ$ dominate the Higgs width, the decays to
gluons provide no real additional suppression of the
$h\to\gamma\gamma$ branching fraction. Finally we note that the
interference of $\CO_G$ with the SM gives simultaneously
larger (smaller) Higgs cross-sections and larger (smaller) $\Gamma(h\to gg)$.

The sensitivity of any experiment to new physics in the Higgs channel
is then a function of several variables:
$m_h$, $f_\gamma$, $f_G$, $\Lambda_\gamma$ and $\Lambda_G$.
There are four sign choices for $f_\gamma, f_G$; we choose to study
the two cases which maximize/minimize the signal at
current and future colliders. The maximum signal case has $f_\gamma=+1$ and
$f_G=-1$; we checked that over the entire range of interest the
increase in the cross-section implied by $f_G=-1$ more than offset the
corresponding increase in $Br(h\to gg)$. The minimum signal case
has the opposite choice of both signs.

Our analysis then has two parts. First we ignore the $\CO_G$ operator
(\ie, $f_G=0$) and
determine the sensitivity of current and future experiments to new physics
through $\CO_\gamma$ alone. In this case, the production cross-section
is simply that of the SM.
Then in a second analysis we include both $\CO_\gamma$ and $\CO_G$.
As we already noted, the effect of $\CO_G$ is both to enhance the production
but also to diminish the relative branching ratio of
$h\to\gamma\gamma$.

For the purposes of doing the numerical calculations, we have used (in
a greatly modified form) the programs of M.~Spira and
collaborators~\cite{spira}. In all cases, we will work only to leading
order. In the SM it has been found that NLO QCD corrections can change
the cross-sections and decay widths by $\sim60\%$~\cite{nlo}. Naively such
changes appear to correspond only to $\sim10\%$ shifts in $\Lambda$, which are
too small for the physics we are interested in here. However, it is
possible that interference effects and enhanced backgrounds (\ie,
$h\to\gamma\gamma$ in the SM) could produce a larger effect ---
we will not consider that possibility here.

Throughout our analysis we also have to address issues of acceptances and
backgrounds in an approximate manner. In Run I, CDF reported an
efficiency times acceptance approaching 15\% in
inclusive $\gamma\gamma+X$ Higgs searches~\cite{cdf}; we will assume
that this figure prevails at all future facilities. There are also two
major sources of backgrounds for our $\gamma\gamma$ signal: SM
processes which produce or fake $\gamma\gamma$, and the usual SM decay of
$h\to\gamma\gamma$ itself. The latter can be calculated explicitly.
For the former we estimate by fitting to the CDF background
spectrum~\cite{cdf}, appropriately scaled to the luminosity of future Tevatron
runs, or the ATLAS background spectrum~\cite{atlas} appropriately scaled for
LHC runs.

\begin{figure}
\centering
\epsfxsize=6truein
\epsffile{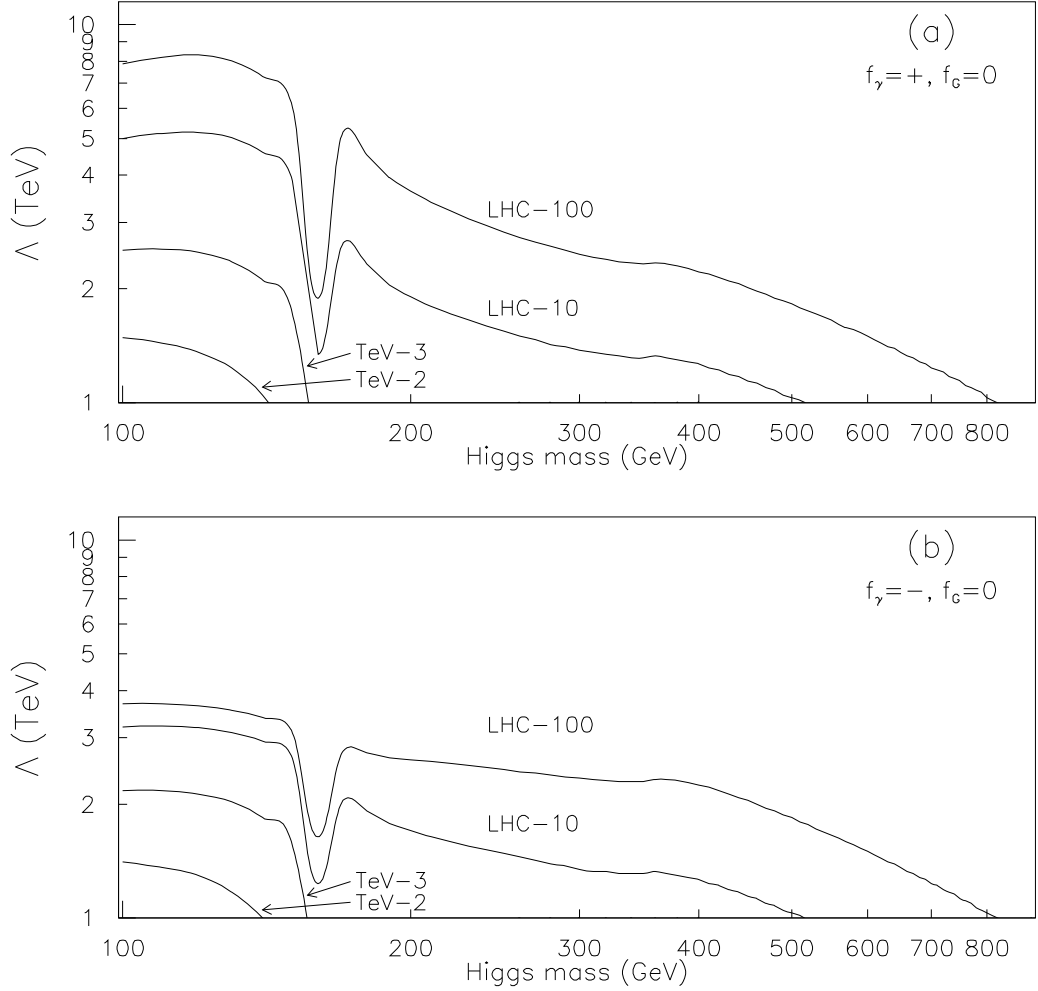}
\caption{$5\sigma$ discovery reaches for $pp,p\bar p\to h\to\gamma\gamma$ in
current and future colliders. Only the $\CO_\gamma$
operator has been included.
In (a), signs are chosen to maximize the signal, while they
are chosen to minimize the signal in (b).}
\label{fig1}
\end{figure}

In Figures~\ref{fig1}(a)-(b)
we show the sensitivity to $\Lambda_\gamma$ that can be
obtained at various machines by plotting their $5\sigma$ discovery reaches
(with no $\CO_G$ contribution). The colliders shown are: the
Tevatron with $\sqrt{s}=1.8\tev$ and $100\invpb$ of luminosity (Run I), with
$\sqrt{s}=2\tev$ and $2\invfb$ of luminosity (Run II), with $\sqrt{s}=2\tev$
and $30\invfb$ (a proposed Run III), and the LHC with $\sqrt{s}=14\tev$ and
$10\invfb$ (initial luminosity) and $100\invfb$ (final luminosity)
respectively. (Note that the TeV Run I line falls below the region of
parameter space plotted.)
As one expects, once the $h\to WW,ZZ$ threshold opens up at $\sqrt{s}\simeq
150\gev$, the large $\Gamma(h\to WW,ZZ)$ is sufficient to overwhelm the
photonic width and our experimental sensitivity drops significantly.
Nonetheless, given the possibility of a light Higgs (and the robust
arguments for one in supersymmetric frameworks) experimentalists
should be encouraged
to view $h\to\gamma\gamma$ as a viable and potentially large signal.

In terms of extracting a conservative discovery reach for $\Lambda$,
Figure~\ref{fig1}(b)
should be used since it chooses $f_\gamma$ in order to minimize
the signal. We note, for example, that the data from Run I cannot
presently probe (or exclude) $\Lambda$ above $1\tev$,
but that Run II should have
a reach of approximately 1 -- $1.5\tev$ for a light Higgs. However it is
important to realize that for generic $f_\gamma$, the various colliders may
have reaches as high as those shown in Figure~\ref{fig1}(a). Thus, for example,
if the Higgs mass is below the $WW$ threshold, the LHC can possibly find a
signal for $\Lambda$ up to $8\tev$ for a light Higgs! (Unfortunately, that
scale could also be as low as $4\tev$.)

\begin{figure}
\centering
\epsfxsize=6truein
\epsffile{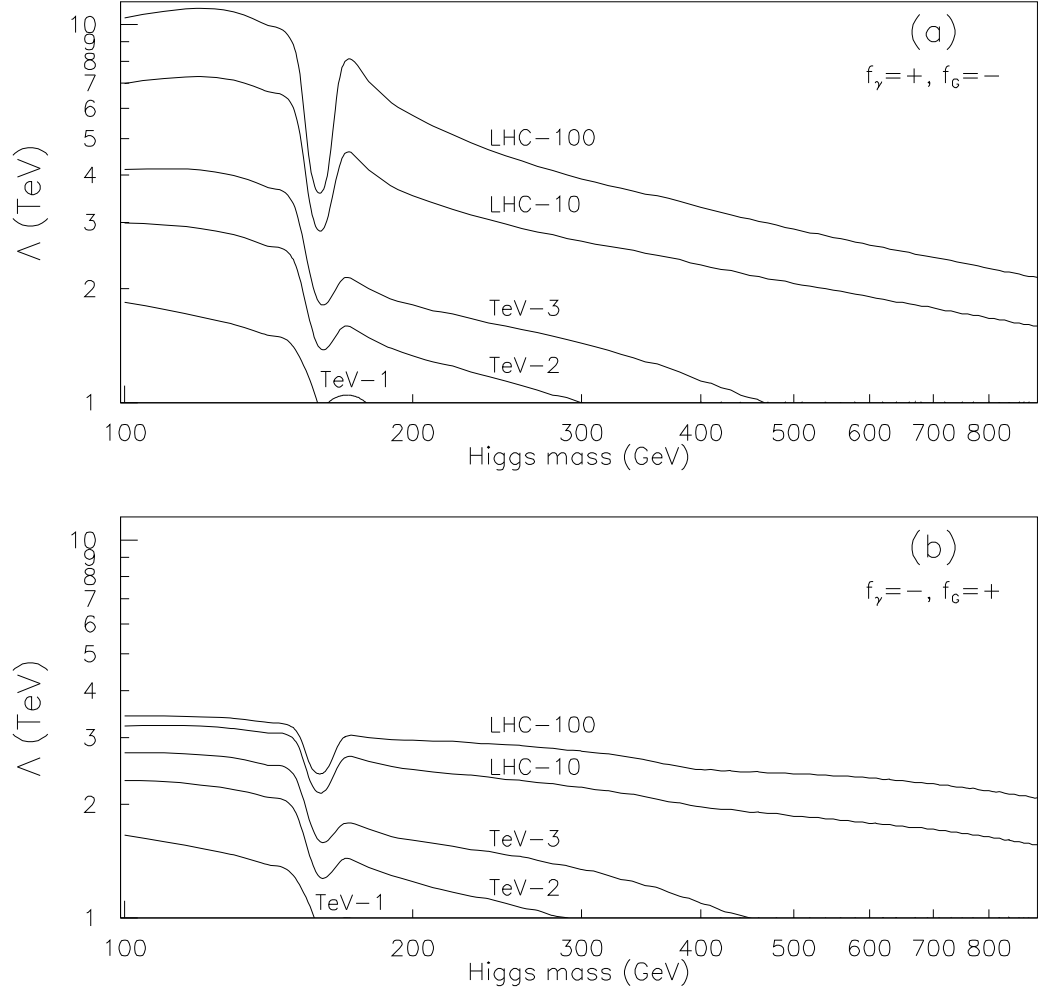}
\caption{$5\sigma$ discovery reaches for $pp,p\bar p\to h\to\gamma\gamma$ in
current and future colliders. Both $\CO_G$ and $\CO_\gamma$ have been
included, with $\Lambda_\gamma=\Lambda_G\equiv\Lambda$.
In (a), signs are chosen to maximize the signal, while they
are chosen to minimize the signal in (b).}
\label{fig2}
\end{figure}

Figures~\ref{fig2}(a)-(b) repeat the same analysis, but now with $\CO_G$
included such that $\Lambda_G=\Lambda_\gamma\equiv\Lambda$.
We view these results
as more realistic compared to those above in which only the
$\CO_\gamma$ operator was kept. We again show the same
set of 5 collider options. Figure~\ref{fig2}(b) is the conservative $5\sigma$
discovery reach, chosen to minimize the $pp,p\bar p\to h\to\gamma\gamma$ rate.
It is interesting
that for a light Higgs, the limits are slightly stronger than those
obtained with $f_G=0$; now even the Tevatron Run I data has the ability to
probe scales above $1\tev$. However the more noticable difference is the
ability to produce larger numbers of heavy Higgs bosons and observe their
$\gamma\gamma$ decays. For example, the LHC is capable of probing scales near
$2\tev$ even for $m_h=1\tev$.

Figure~\ref{fig2}(b) shows the maximal reach of the various
colliders, with the LHC now extending its sensitivity to $\Lambda$
as high as $10\tev$ for a light Higgs! Finally, we summarize a few of our
results for $m_h=110$, $200$ and $500\gev$ for both exclusion and discovery in
Table~\ref{table1}. All bounds assume $\Lambda_\gamma=\Lambda_G$. For each
choice of the Higgs mass, we have shown a conservative limit on $\Lambda$ which
can be excluded, and a maximum $\Lambda$ below which a signal may be
discovered. Thus for
the exclusion bounds ($2\sigma$) we have taken the interference effects to
minimize the signal; for the maximum discovery reaches ($5\sigma$), we have
chosen the interference effects to maximize the signal.
\begin{table}
\centering
\begin{tabular}{l|cc|cc|cc}
\multicolumn{1}{c}{~} & \multicolumn{6}{c}{$m_h$ (GeV)} \\
\multicolumn{1}{c|}{$\Lambda$ (TeV)} & \multicolumn{2}{c}{110} &
\multicolumn{2}{c}{200} & \multicolumn{2}{c}{500} \\ \hline
Tev Run I & 2.0 & 1.8 & 1.1 & --- & --- & --- \\
Tev Run II & 2.6 & 3.0 & 1.5 & 1.3 & --- & --- \\
Tev Run III & 3.0 & 4.2 & 1.8 & 1.8 & 1.1 & --- \\
LHC ($10\invfb$) & 3.4 & 7.2 & 2.9 & 3.5 & 2.3 & 2.1 \\
LHC ($100\invfb$) & 3.5 & 10.8& 3.2 & 5.8 & 2.9 & 2.9 \\
\end{tabular}
\label{table1}
\caption{Exclusion limits and maximum discovery reaches (in TeV) for various
collider runs for 3 representative Higgs masses. The first column for each
$m_h$ is a conservative $2\sigma$ exclusion reach for each machine; the second
column is the optimistic $5\sigma$ discovery reach. Unfilled columns represent
limits below $1\tev$. We take $\Lambda_G=\Lambda_\gamma$ for the table.}
\end{table}

We have attempted in this analysis to be rather conservative. For one thing,
the $2\sigma$ exclusion limits of the various colliders are often several TeV
higher than the $5\sigma$ discovery limits. Secondly, we have treated the
discovery of the $h\to\gamma\gamma$ signal as simply a counting experiment,
throwing away useful experimental information, for example on the shape of the
diphoton mass spectrum, which would be available experimentally to help extract
the signal from the backgrounds. Lastly, we have not included QCD corrections
to the amplitudes, which we believe could increase the signal (though also
increasing the ``background'' $h\to\gamma\gamma$ signal) by $\sim50\%$.
Therefore we believe that the reaches given here are to be taken as
conservative values, insofar as one should take the scales deduced from naive
power-counting seriously.

\section{Conclusions}

In this paper we have studied two consequences of large extra
dimensions for electroweak symmetry breaking: a relaxation of the precision
electroweak bound on the Higgs boson mass, and an enhanced rate for
$\gamma \gamma$ events at hadron colliders from Higgs decay.

The relaxation of the precision electroweak bound on the Higgs mass
applies when any new physics generates (\ref{obw}) and (\ref{op1}) at
a scale of several TeV. It is well known that
$S$ and $T$ depend only logarithmically on the Higgs boson mass,
but it may not be appreciated that the mass bound can be
evaded completely for a wide range of values of $\Lambda$, extending
as high as 10 TeV. For
example, even a weakened bound of $m_h < 500 \gev$, only applies if 
the standard model is
the correct description of nature up to energies of $17\tev$.
We find this implausible, since it implies a fine tuning
in the Higgs mass squared parameter of 1 part in 2000. There is only one
strong argument for a light Higgs boson: the correct successful
prediction of the weak mixing angle at the percent level of accuracy
requires weak scale supersymmetry, and therefore a light Higgs
boson. In theories with large extra dimensions this argument is not
applicable, since the percent level prediction for the weak mixing angle
is lost. Hence, in these theories, there is no preference for a
light Higgs boson, and thus
alternatives with a heavy Higgs or no Higgs should be considered seriously.

If there is a Higgs boson, we have shown that a generic signal of
large extra dimensions is an anomalously large
$\gamma\gamma$ signal at machines capable of producing Higgs bosons. 
Expectations from the SM put such a signal out of reach of the
Tevatron.  In Figure~\ref{fig2}
we showed the $5 \sigma$ discovery reaches for $h
\rightarrow \gamma \gamma$ at the Tevatron and LHC. At Run II of the
Tevatron collider this signal would be discovered for a light Higgs if
$\Lambda$ is less than 2 (3) TeV for destructive (constructive) interference.
LHC not only increases the discovery potential for a light Higgs boson 
mass, up to $10\tev$ for constructive interference, but also has significant
discovery potential up to the largest Higgs masses. This signal
compares favorably with that of graviton production at colliders
\cite{graviton}, especially if the scale which sets the size of the $4+n$
dimensional gravitational coupling is somewhat larger than the
scale $\Lambda$.

\section*{Acknowledgements}
We are grateful to Nima Arkani-Hamed, Michael Chanowitz, Savas Dimopoulos and
Henry Frisch for many
useful conversations. This work was supported in part by the U.S.\
Department of Energy under contract DE--AC03--76SF00098 and by the
National Science Foundation under grant PHY--95--14797.


\begin{thebibliography}{99}

\bibitem{lowmpl}
      N.~Arkani-Hamed, S.~Dimopoulos and G.~Dvali, \PLB{429}{98}{263} and
      \PRD{59}{99}{086004};\\
      N.~Arkani-Hamed, S.~Dimopoulos and J.~March-Russell,
      {\tt hep-th/9809124.}

\bibitem{bdn}
      T.~Banks, M.~Dine and A.~Nelson, {\tt hep-th/9903019}.

\bibitem{einhorn}
      C.~Arzt, M.~Einhorn and J.~Wudka, \NPB{433}{95}{41}.

\bibitem{hagiwara}
      K.~Hagiwara, R.~Szalapski and D.~Zeppenfeld,
      \PLB{318}{93}{155};\\
      M.~Gonzalez-Garcia, S.~Lietti and S.~Novaes,
      \PRD{57}{98}{7045}.

\bibitem{apv}
      A.~Deandrea, \PLB{409}{97}{277}.

\bibitem{eebb}
      K.~Ackerstaff \etal\ (OPAL Collaboration), \PLB{391}{97}{221}.
      An unpublished result presented by R.~Clare at the
      LEPC meeting of November 1998 is only slightly stronger.

\bibitem{mumuqq}
      F.~Abe \etal\ (CDF Collaboration), \PRL{79}{97}{2198}.

\bibitem{hagiold}
      K.~Hagiwara, T.~Hatsukano, S.~Ishihara and R.~Szalapski,
      \NPB{496}{97}{66}; \\
      K.~Hagiwara, D.~Haidt and S.~Matsumoto,
      Eur.\ Phys.\ J.\ {\bf C2} (1997) 95.

\bibitem{erler}
       J.~Erler, {\tt hep-ph/9903449}.

\bibitem{kkw}
      G.~Kane, C.~Kolda and J.~Wells, \PRL{70}{93}{2686};\\
      J.~Espinosa and M.~Quiros, \PLB{302}{93}{51}.

\bibitem{ad} N.~Arkani-Hamed and S.~Dimopoulos, {\tt hep-ph/9811353}.

\bibitem{nlo}
      M.~Spira, A.~Djouadi, D.~Graudenz and
      P.~Zerwas, \NPB{453}{95}{17}.

\bibitem{eboli}
      O.~\'Eboli, M.~Gonzalez-Garcia, S.~Lietti and S.~Novaes,
      \PLB{434}{98}{340}.

\bibitem{spira}
      M.~Spira, {\tt hep-ph/9510347}; \\
      A.~Djouadi, J.~Kalinowski and M.~Spira, {\tt hep-ph/9704448}.

\bibitem{cdf}
      P.~Wilson (CDF Collaboration), ``Search for High Mass Photon
      Pairs in $p\bar p$ Collisions at $\sqrt{s}=1.8\tev$,'' contributed
      to ICHEP 98, preprint FERMILAB-CONF-98/213-E (Jun 1998).

\bibitem{atlas}
      W.~Armstrong \etal\ (ATLAS Collaboration), ``ATLAS Technical
      Proposal,'' preprint CERN/LHCC/94-43 (Dec 1994).

\bibitem{graviton}
      G.~Giudice, R.~Rattazzi and J.~Wells, {\tt hep-ph/9811291};  \\
      E.~Mirabelli, M.~Perelstein and M.~Peskin, {\tt hep-ph/9811337}.

\end{thebibliography}
\end{document}